\documentclass{appolb} \usepackage{epsfig}
\begin{document}

\title{Strong Deformation Effects in Hot Rotating $^{46}$Ti\thanks{
       Presented at Zakopane Conference on Nuclear Physics,  
       September 4-10, 2006}
       }

\author{M.~Kmiecik$^a$, A.~Maj$^a$, M.~Brekiesz$^a$, K.~Mazurek$^{a}$,
        P.~Bednarczyk$^{a}$, J.~Gr\c{e}bosz$^a$, W.~M\c{e}czy\'nski$^a$,
        J.~Stycze\'n$^a$, M.~Zi\c{e}bli\'nski$^a$, K.~Zuber$^a$, 
        P.~Papka$^{b,c}$, C.~Beck$^c$, D.~Curien$^c$, F.~Haas$^c$, V.~Rauch$^c$,
        M.~Rousseau$^c$, J.~Dudek$^c$, N.~Schunck$^d$,
        A.~Bracco$^e$, F.~Camera$^e$, G.~Benzoni$^e$,  O.~Wieland$^e$, 
        B.~Herskind$^f$, E. Farnea$^g$, G. De Angelis$^h$\\  
        \address{$^a$The Niewodnicza\'nski Institute of Nuclear Physics, PAN,
        Krak\'ow, Poland \\
        $^b$iThemba LABS, Somerset West, South Africa\\
        $^c$IPHC IN$_{2}$P$_{3}$-CNRS/Universit\'e Louis Pasteur, 
        F-67037 Strasbourg Cedex 2, France\\ 
        $^d$Depart\'amento de F\'isica Te\'orica, Universidad Aut\'onoma de
        Madrid, Spain\\
        $^e$University of Milano - Department of Physics and INFN section of
         Milano, Italy\\
        $^f$The Niels Bohr Insitute, Copenhagen, Denmark\\
        $^g$INFN sez. Padova, I-35131 Padova, Italy\\
        $^h$INFN - Laboratori Nazionali di Legnaro, I-35020 Legnaro (PD), Italy
}
}

\maketitle
 
\begin{abstract}
  
  Exotic-deformation effects in $^{46}$Ti nucleus were investigated 
  by analysing the high-energy $\gamma$-ray and the $\alpha$-particle energy
  spectra. One of the experiments was performed using the charged-particle
  multi-detector array ICARE together with a large volume (4"$\times$4") BGO
  detector. The study focused on simultaneous measurement of light charged
  particles and $\gamma$-rays in coincidence with the evaporation residues.  
  The experimental data show a signature of very large deformations 
  of the compound nucleus in the Jacobi transition region at the highest spins. 
  These results are compared to data from previous experiments performed 
  with the HECTOR array coupled to the EUROBALL array, where it was found 
  that the GDR strength function is highly fragmented, strongly indicating 
  a presence of nuclei with very large deformation.

\end{abstract}

  
\section{Introduction}

Theoretical calculations using the recent Lublin-Strasbourg Drop (LSD) 
model~\cite{Pom03,Dub05}, predict the Jacobi shape transitions in $^{46}$Ti 
in the spin region of $I \sim$~30-40~$\hbar$ implying an existence of the dramatic
shape instability and a rapid increase in elongation corresponding to relatively
small spin changes. More precisely, the nucleus changes its shape from an oblate
- with the spin parallel  to the symmetry axis - to an elongated prolate or
triaxial shape at high spins. The results of the LSD model calculations are
shown in Fig.~1, where the potential energy distributions of $^{46}$Ti 
are plotted in the $(\beta,\gamma)$-plane for six selected values of spin. 
The minimum (shown with a~circle) corresponds to the equilibrium shape 
of the nucleus at each given spin. The Figure illustrates the energy-minimum 
evolution with increasing spin - the evolution path is shown in the Figure 
with solid line. The nucleus is nearly spherical at low angular momenta; 
it increases its oblate deformation up to $I\approx$~26~$\hbar$ 
where it becomes triaxial. For the highest spins, the nucleus becomes 
strongly elongated, it becomes nearly axial at $I \approx$~38~$\hbar$ 
where the fission barrier amounts to $\sim$~6~MeV only.

\begin{figure}[htb]
\begin{center}
\epsfig{file=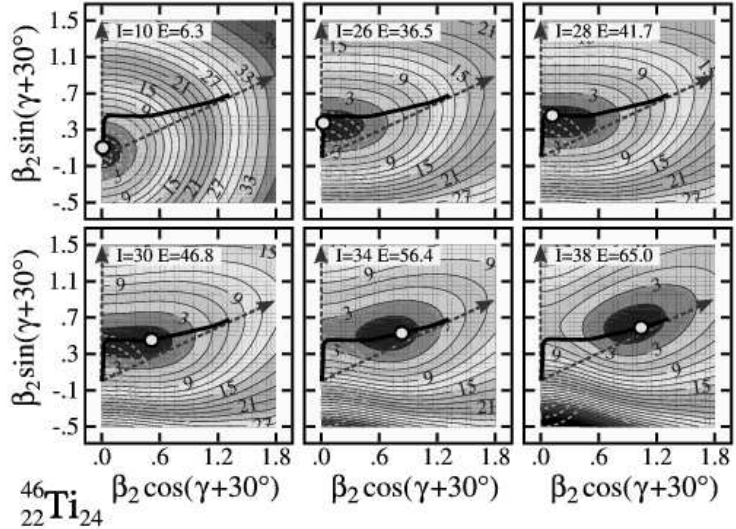,height=7.0cm}
\caption{Potential energies for $^{46}$Ti nucleus as a functions of quadrupole
         $\beta_2$ and $\gamma$ deformations. At each $(\beta_2,\gamma)$-point
         the energy was minimised over $\beta_{40}$, $\beta_{60}$ 
         and $\beta_{80}$. The corresponding spin values $I$ are given 
         in every panel together with the calculated minimum energies $E$. 
}
\end{center}
\end{figure}

The large deformation effects in the light-mass nuclei have been studied over 
the last years, using both gamma and charged-particle spectroscopy. In particular 
the very elongated prolate or triaxial shapes were observed  from the spectra 
of the Giant Dipole Resonance (GDR) decay  for $^{46}$Ti$^*$~\cite{Maj04} 
and $^{45}$Sc$^*$~\cite{Kic93}. Large deformations were also observed 
in $^{44}$Ti$^*$~\cite{Pap03} by the measurement of light charged particle (LCP) 
spectra originating from the statistical decay of this compound system. 
Additionally, in this mass region, a number of superdeformed bands of discrete  
$\gamma$-ray transitions were discovered (cf. e.g.~\cite{Lach02,Ide01,CBeck}).

%%%%%%%%%%%%%%%%%%%%%%%%%%%%%%%%%%%%%%%%%%%%%%%%%%%%%%%%%%%%%%%%%%%%%%%%%%%%%%%%
%%%%%%%%%%%%%%%%%%%%%%%%%%%%%%%%%%%%%%%%%%%%%%%%%%%%%%%%%%%%%%%%%%%%%%%%%%%%%%%%

\section{The GDR from the Highest Spin Region of the $^{46}$Ti Nucleus}

The shape of the $^{46}$Ti nucleus at high spins was first studied 
in the experiment~\cite{Maj04} performed using HECTOR BaF$_2$ detectors~\cite{Maj94}
coupled to the EUROBALL~IV HPGe detector array. The excitation energy 
of the $^{46}$Ti nuclei, populated in the $^{18}$O+$^{28}$Si reaction 
at 105~MeV bombarding energy was $E^*$~=~85~MeV and the maximum angular momentum, 
$L_{max}\approx$~35~$\hbar$. High energy $\gamma$-rays 
coming from the GDR decay (detected in HECTOR) were measured 
in coincidence with the gamma multiplicity measured in the Innerball,
and with the discrete transitions in the $^{42}$Ca final nucleus 
identified with the help of EUROBALL.
\begin{figure}[htb]
\begin{center}
\epsfig{file=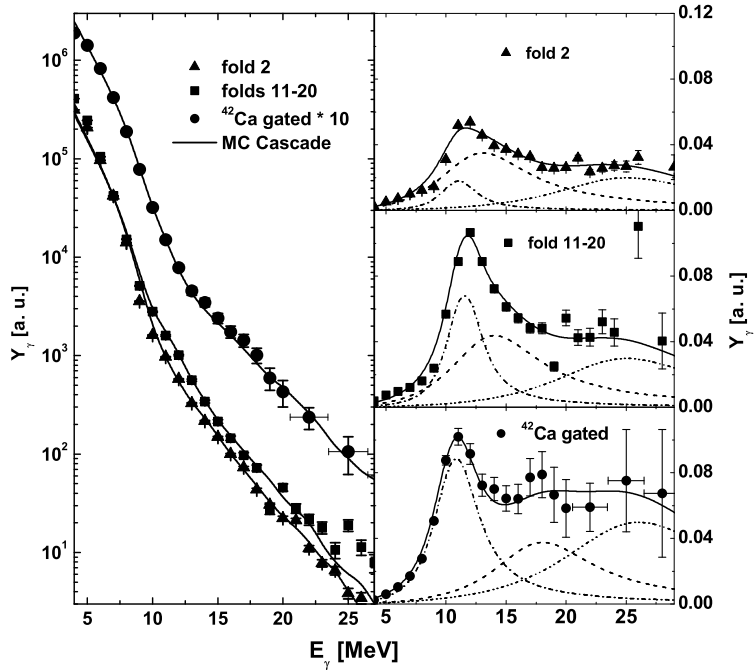,height=9.0cm}
\caption{{\it Left panel:} The high energy spectra measured 
        for two regions of fold corresponding to high and low spins 
        shown together with a spectrum gated with discrete transitions 
        in $^{42}$Ca. {\it Right panel:} The GDR strength function 
        obtained from Monte Carlo Cascade fit to the experimental data.
}
\end{center}
\end{figure}
We have analyzed the GDR spectra for different gamma folds (defined by the
number of Innerball detectors that fired) corresponding to various spin regions.
The spectra for fold~=~2 and folds~=~11-20 are presented in the left-hand side
panel of Fig.~2 together with the high energy gamma spectrum measured in the
decay channel leading to $^{42}$Ca, selected by requiring coincidences with
discrete transitions in this final nucleus.

To compare the GDR line-shapes (GDR strength functions)
 all experimental spectra were linearized,
 using the method described in~\cite{Kic93}, by fitting 
the experimental spectra with the spectra calculated by the Monte Carlo 
version of the statistical 
model code CASCADE~\cite{Puh77}.
 The GDR strength functions, obtained 
from the fit to the experimental data measured for different folds 
(the top and middle right-hand panel of Fig.~2), 
show a fragmented structure with a low energy component growing with spin 
together with two broad components at significantly higher energies. 
Similar line-shape of the GDR (the bottom right-hand panel of Fig.~2),
 but with an even larger splitting,
was obtained for the high energy spectra gated by the discrete transitions 
in $^{42}$Ca~\cite{Maj04}.  
It was demonstrated (see~\cite{Maj04}) that choosing this decay channel 
we select the Ti compound nuclei of the highest spins such 
that the splitting of the GDR strength is the largest possible, as shown 
in Fig.~2. 

\begin{figure}[htb]
\begin{center}
\epsfig{file=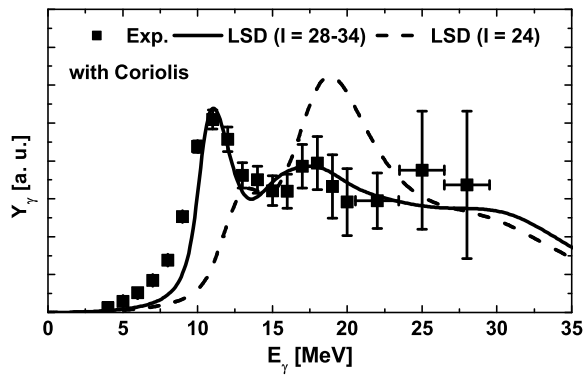,height=5.cm}
\caption{The experimental GDR strength function obtained in Ref.~\cite{Maj04}
         from data gated by discrete transitions in $^{42}$Ca (points) 
         compared to LSD model calculations obtained for two spin regions 
         shown by dashed line ($I$~=~24~$\hbar$ ) and solid line 
         ($I$~=~28-34~$\hbar$).
}
\end{center}
\end{figure}

In Figure~3 the experimental $^{42}$Ca gated strength function is shown 
again and compared to the LSD model calculations for two different 
spin regions: $I~\leq~24~\hbar$ -~governed mostly by oblate shapes 
and $I$~=~28-34~$\hbar$ pointing to strongly elongated prolate shapes 
($\beta \approx$~0.8) in the upper Jacobi transition region just below 
the fission limit~\cite{Bec96}. One should note that the limiting 
angular momentum for fusion is predicted by the Finite-Range Liquid Drop 
Model (FRLDM), consistent with LSD~\cite{Pom03,Dub05}), to be around 
35~$\hbar$ in agreement with experimental data~\cite{Bec96}.
A significant Coriolis splitting at high spin was included in these 
calculations similarly as in Refs.~\cite{Dub05,Maj04}.

The GDR line-shape calculated for the Jacobi shape transition region  
is in good agreement with the experimental data proving the existence  
and importance of the Jacobi shape transition and the strong Coriolis effect. 
Our estimates give the splitting of the two GDR components 
by $\Delta E\approx$~5~MeV. Such strong Coriolis effect on a GDR spectrum 
was observed for the first time in the experiment presented in Ref.~\cite{Maj04}.

%%%%%%%%%%%%%%%%%%%%%%%%%%%%%%%%%%%%%%%%%%%%%%%%%%%%%%%%%%%%%%%%%%%%%%%%%%%%%%%%
%%%%%%%%%%%%%%%%%%%%%%%%%%%%%%%%%%%%%%%%%%%%%%%%%%%%%%%%%%%%%%%%%%%%%%%%%%%%%%%%

\section{Deformation Studied by the Charged-Particle Spectra}

The very large deformations presented in the previous section for $^{46}$Ti  
at high angular momentum, are also suggested in the following by the present 
study of the $\alpha$-particle spectra measured in the experiment performed 
at the Strasbourg Vivitron tandem facility using the multi-detector array
ICARE~\cite{Pap03,Rou02} together with a large volume (4"$\times\,$4") BGO
detector. The compound nucleus $^{46}$Ti was populated at the excitation energy
of $E^*$=~85~MeV and at angular momenta approaching $L_{max}\approx$~35~$\hbar$.
The latter angular momentum limit was similar to the one reached 
in the experiment discussed above; it is close to the fission limit  predicted 
by the FRLDM~\cite{Bec96}, except for the inverse kinematics reaction 
corresponding to 144~MeV $^{27}$Al beam on $^{19}$F target. The heavy fragments 
were detected in 10 telescopes, each consisting of an ionisation chamber (IC) 
followed by a 500~$\mu$m Si detector, located at $\Theta$$_{Lab}$~=~$\pm$10$^{\circ}$  
in three reaction planes. The light charged particles were measured using 10 triple
telescopes (40~$\mu$m Si, 300~$\mu$m Si, 2~cm CsI(Tl)) and 18 two-element
telescopes (40~$\mu$m Si, 2~cm CsI(Tl))~\cite{Bre05,Bre06}.

\begin{figure}[h]
\begin{center}
\epsfig{file=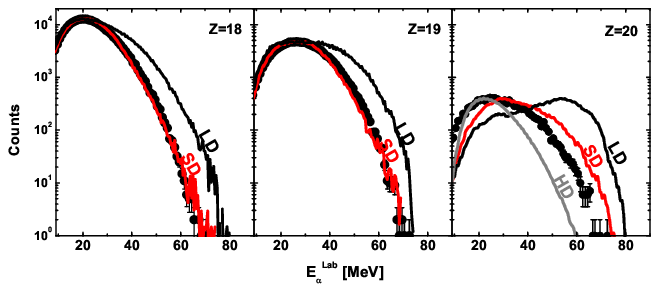,height=5.0cm}
\epsfig{file=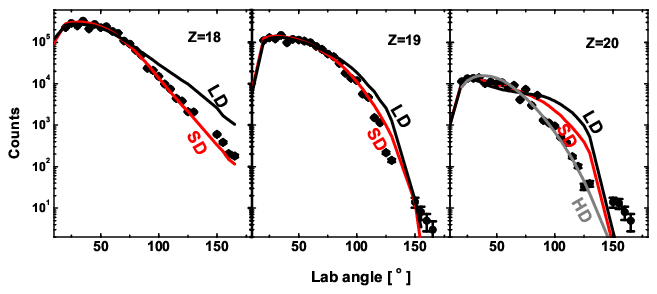,height=5.0cm}
\caption{{\it Top panel:} The $\alpha$-particle experimental spectra measured 
        at $\Theta _{Lab}=$~45$^{\circ}$ in laboratory frame 
        (from Ref.~\cite{Bre06}). 
        {\it Bottom panel:} The $\alpha$-particle angular correlations 
        measured in laboratory frame in coincidence with residues detected 
        at $\Theta _{Lab}=~-$10$^{\circ}$. The lines presented in both panels 
        are statistical model calculations performed for different deformation
        parameters as explained in the text.}
\end{center}
\end{figure}

The energy spectra of the $\alpha$-particles emitted in the laboratory frame 
at the angle $\Theta _{Lab}=$~45$^{\circ}$ in coincidence with the residual 
nuclei of $Z$~=~18,~19 and 20 are shown in the top panel of Fig.~4 by solid points. 
The bottom panel presents the angular correlations of the $\alpha$-particles
with the evaporation residues (ER) of $Z$~=~18,~19 and 20  detected in the IC
placed  at $\Theta _{Lab}=~-$10$^{\circ}$. The lines are the results  
of the analysis performed using the code CACARIZO~\cite{Mah04}, 
the LCP Monte Carlo version of the statistical model code CASCADE~\cite{Puh77}, 
for several hypotheses concerning the yrast line.

The high energy part of the $\alpha$-particle spectra depends on the final state
level density. The level density is calculated in the code using  the Rotating
Liquid Drop Model (RLDM)~\cite{Puh77} and can be changed  using different sets
of deformation parameters describing the yrast line  in question. In the code,
the yrast line is parameterized by the numerical values of  $r_0$, $\delta$$_1$
and $\delta$$_2$ and described by the formula: 
$E_L = \hbar^2L(L + 1)/2\Im_{eff}$ where $\Im_{eff}$ denotes the effective 
moment of inertia defined as $\Im_{eff}=\Im_{sphere}(1+\delta_1L^2+\delta_2L^4)$  
and $\Im_{sphere}=(2/5)A^{5/3}r_0^2$ is the rigid body moment of inertia   
of the spherical nucleus; $r_0$ is the radius parameter (see Ref.~\cite{Mah04} 
for more details).

In Fig.~5, the yrast lines used in the CACARIZO calculations are displayed 
as solid lines. The standard RLDM yrast line (shown as ``liquid drop") 
can be approximated by the rigid body yrast line with small deformation
($\beta$~=~0.2). The calculated $\alpha$-particle spectra and angular 
correlations obtained for this parameterization, denoted as ``LD", 
do not reproduce the experimental spectra.

\begin{figure}[htb]
\begin{center}
\epsfig{file=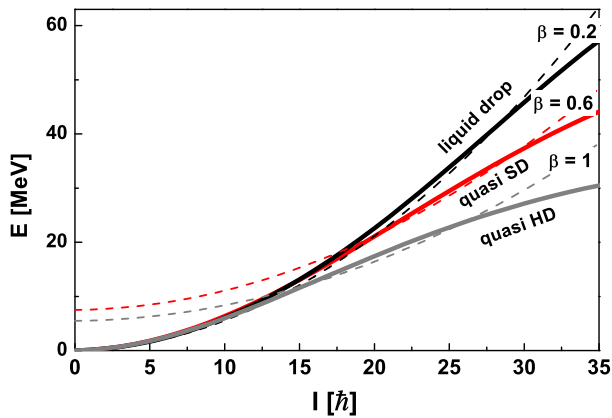,height=5.5cm}
\caption{The yrast lines implemented in the calculations (from
         Ref.~\cite{Bre06}). 
}
\end{center}
\end{figure}

The yrast line, denoted quasi-superdeformed (``quasi SD"), corresponds 
to the spin region $I$~=~15-30~$\hbar$ for the yrast line of the rigid body 
with deformation parameter $\beta \approx$~0.6. Actually, the yrast line 
with the same deformation parameters was deduced from a reasonable description 
of the $\alpha$-particle decay of $^{44}$Ti~\cite{Pap03}. The calculations 
with this parameterization, shown as ``SD", reproduce the experimental spectra 
for $Z$~=~18 and $Z$~=~19 while the spectra associated with $Z$~=~20 still deviate
significantly from the model results. In order to improve the agreement 
for $Z$~=~20 we have been forced to assume an even more deformed yrast line. 
In this calculation, the yrast line labeled  in Fig.~5 ``quasi HD" was used 
[this label was used as the corresponding line  resembles the yrast line 
for the rigid body with a deformation parameter  
$\beta\approx$~1 (for $I$~=~15-30~$\hbar$)]. However, the ``HD" calculations 
(shown in Fig.~4) underestimate the experimental data of energy distributions 
in this case,  pointing to deformations between $\beta=$~0.6 and 1, 
while the simulated angular correlations agree completely with the data.

%%%%%%%%%%%%%%%%%%%%%%%%%%%%%%%%%%%%%%%%%%%%%%%%%%%%%%%%%%%%%%%%%%%%%%%%%%%%%%%%
%%%%%%%%%%%%%%%%%%%%%%%%%%%%%%%%%%%%%%%%%%%%%%%%%%%%%%%%%%%%%%%%%%%%%%%%%%%%%%%%

\section{Discussion of the Results}

A possible explanations of such an anomalously large deformation, that may
seem unrealistic, can be related to the time scale of the evaporation process.

When many particles are evaporated, the time needed for this process 
can be long enough, such that it is sufficient for the residual nucleus 
to adjust its initial ''Jacobi shape" (with $\beta \approx$~1)
to smaller deformations ($\beta \approx$~0.6) at lower temperatures and spins.
The effective level density of the final states has to be described 
by a different deformation as for the initial Jacobi shapes, i.e. 
by the quasi-superdeformed yrast line.

However, for $Z$~=~20, with only a single $\alpha$-particle emission, 
the process time may be too short to change the shape. Therefore the yrast line 
describing such nucleus may lie between the quasi-hyperdeformed yrast line 
(with $\beta \approx$~1), describing the initial Jacobi shapes,
and a quasi-superdeformed yrast line (with $\beta \approx$~0.6), 
describing the final states. The deformation of a nucleus visualised 
in the spectra of $\alpha$-particles emitted during the process 
of changing the shape of a nucleus may be considered 
to be {\it ``dynamical hyperdeformation"}. Similar result 
was in fact observed in the $\alpha$-particle spectra from the decay
of $^{59}$Cu~\cite{For89}.

Another explanation related to the evaporation time scale may be considered, 
inspired by the angular distributions of charged particles measured 
by the EUCLIDES array during the HECTOR + EUROBALL experiment~\cite{Maj04}.

\begin{figure}[h]
\begin{center}
\epsfig{file=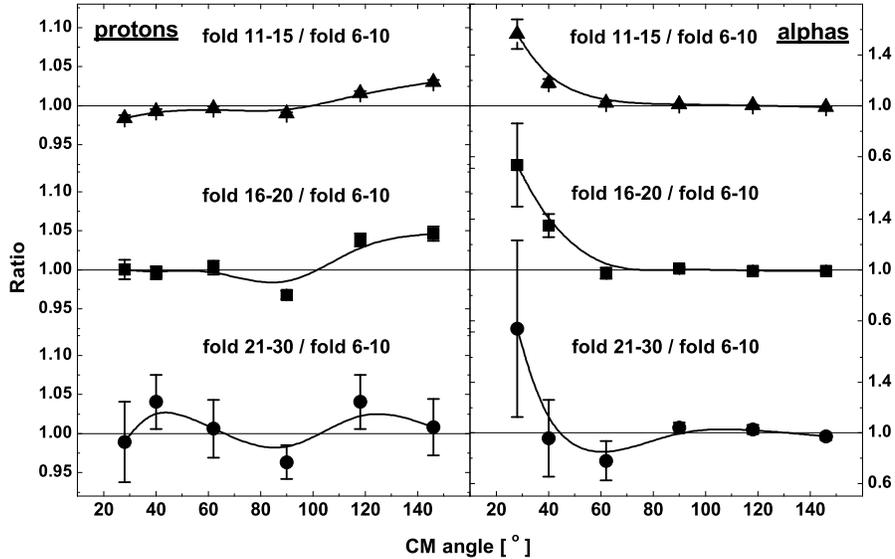,height=7.5cm}
\caption{The angular distributions of protons and $\alpha$-particles, 
         measured using the EUCLIDES array. The experimental data are
         represented with points. The solid lines are to guide the eye.
}
\end{center}
\end{figure}

The angular distributions measured in the laboratory frame for different 
fold windows were normalized to the one measured at fold region 5-10, 
and converted to the center of mass.
Fig.~6 shows such relative angular distributions of protons and $\alpha$-particles
for different fold regions, obtained in coincidence with the low energy 
transitions in $^{42}$Ca simultaneously measured in EUROBALL Ge-array. 
In the case of protons one can see that for the highest folds 
(highest average spin) the distribution becomes symmetric around 90$^{\circ}$, 
as expected for evaporation from the collectively rotating, deformed nucleus.

In contrast, for the $\alpha$-particles the angular distributions obtained are
not symmetric and show forward peak at high spins. This suggests 
a {\it pre-equilibrium emission} of $\alpha$-particles, or perhaps even 
an in-complete fusion process. Such equilibration process, taking place 
especially for a mass-symmetric reaction and at highest spins, is usually 
characterized by extended, ``di-nuclear" systems and by relatively long times 
(up to 10$^{-19}$~s)~\cite{Maj99}. If an $\alpha$-particle is emitted during 
this equilibration process, its emission may in a natural way be described 
by the large deformations (with $\beta \approx$~1) of the ``di-nuclear" systems. 
Similar effect may also have been observed in the much heavier symmetric fusion 
reaction $^{64}$Ni~+~$^{64}$Ni leading to hyperdeformed quasi-continuum states 
in the $A~=$~125 region~\cite{Her06}.

%%%%%%%%%%%%%%%%%%%%%%%%%%%%%%%%%%%%%%%%%%%%%%%%%%%%%%%%%%%%%%%%%%%%%%%%%%%%%%%%
%%%%%%%%%%%%%%%%%%%%%%%%%%%%%%%%%%%%%%%%%%%%%%%%%%%%%%%%%%%%%%%%%%%%%%%%%%%%%%%%

\section{Summary}

The deformation effects in $^{46}$Ti were investigated in high energy 
$\gamma$-ray GDR decay as well as with the $\alpha$-particle energy 
and angular distribution measurements. All of them show large deformation
of the $^{46}$Ti compound nucleus at high spins as predicted by the theoretical 
calculations of Fig.~1 performed within the LSD model~\cite{Pom03,Dub05}
and consistent with the SD bands recently discovered in this light-mass
region~\cite{Lach02,Ide01,CBeck}.

The obtained angular distributions confirm the results of the measurements 
of the $\alpha$-particle energy spectra. The $\alpha$-particles emitted 
from hot rotating $^{46}$Ti compound nuclei point to large deformations 
($\beta \approx$~0.6) involved in the evaporation process. In the case 
of a single $\alpha$-particle emission an even larger deformation 
($\beta \approx$~1) is suggested. This may be interpreted as an effect 
of {\it dynamical hyperdeformation} or {\it pre-equilibrium emission} 
of an $\alpha$-particle from a ``di-nuclear system".

The high energy $\gamma$-ray spectra measured in the HECTOR + EUROBALL
experiment definitely show a strongly fragmented structure with the low 
energy component growing with increasing spin and a broad part at higher 
energies. Such line-shape of the GDR corresponds to the expected Jacobi 
shape transition including the enhanced splitting by the Coriolis 
interaction at high spin. In addition, it was found that the low energy 
GDR component seems to feed preferentially the superdeformed band 
in $^{42}$Ca~\cite{Maj05,Kmi05}. This suggests that the very deformed 
shapes after the Jacobi shape transition in the hot compound nucleus 
remain during the evaporation process, especially if it proceeds 
via the single $\alpha$-particle emission channel, feeding $^{42}$Ca.

Clearly, the Jacobi shape transition in the compound nucleus plays 
an important role in population of very elongated rapidly rotating 
cold nuclei, as was proposed in~\cite{Dud05}.

\bigskip

We would like to thank the Vivitron staff, J.~Devin, C.~Fuchs and M.A.~Saettel 
for the excellent support in carrying the experiment at Strasbourg.
This work was partially supported by the Polish Committee for
Scientific Research (KBN Grant No. 1 P03B 030 30), by the exchange
programme between the {\em Institut National de Physique Nucl\'eaire
et de Physique des Particules, $IN_2P_3$} and the Polish Nuclear
Physics Laboratories, the Danish Science Foundation 
and by the European Commission contract HPRI-CT-1999-00078.

\end{document}